\documentclass[pdflatex,sn-vancouver-num]{sn-jnl}

\usepackage[T1]{fontenc}
\usepackage[utf8]{inputenc}
\usepackage{amsmath}
\usepackage{amssymb}
\usepackage{booktabs}
\usepackage{graphicx}
\usepackage{microtype}
\usepackage{siunitx}
\usepackage{xcolor}

\hypersetup{hidelinks}

\graphicspath{{figures/}}
\newcommand{\fig}[1]{Fig.~\ref{#1}}

\begin{document}

\title{Spatial patterns of transportation choice in Niter\'oi}

\author*[1, 2]{\fnm{Sandro M.} \sur{Reia}}\email{smarti71@gmu.edu}
\author[1, 2]{\fnm{Taylor} \sur{Anderson}}
\author[1, 2]{\fnm{Dieter} \sur{Pfoser}}

\affil[1]{\orgdiv{Department of Geography and Geoinformation Science}, \orgname{George Mason University}, \orgaddress{\city{Fairfax}, \country{United States}}}
\affil[2]{\orgdiv{Center for Geospatial Intelligence}, \orgname{George Mason University}, \orgaddress{\city{Fairfax}, \country{United States}}}

\abstract{
Characterizing how people move through a city requires spatially and temporally fine-grained data that are often costly and difficult to obtain.
Here, we introduce a computational framework that uses macroscopic, aggregated, and static urban data to construct a microscopic, dynamic, and spatially detailed representation of urban mobility.
Building on an existing activity simulator, the framework combines a synthetic population, the built environment, and transport networks to generate daily trips and assign walking, car, or public transit.
Applied to approximately 475,000 synthetic residents of Niter\'oi, Brazil, simulated neighborhood-level transit patterns show a strong association with an empirical activity proxy (Pearson $r=0.885$).
Walking is more prevalent in compact central areas, car use increases toward the periphery, and public-transit use remains comparatively high along bus corridors.
Reducing walking times to and from bus stops by 30\% increases the modeled public-transit-to-car ratio by 11.3\%, with larger transit-share gains in several peripheral areas.
These results reveal spatial differences in transportation choice and identify where improved transit access may have the greatest modeled effect.
}


\maketitle

\section{Introduction}
\label{sec:introduction}

Urban mobility emerges from the activities people perform, the places they
visit, and their choices of how to travel between them. Understanding how
these choices vary across neighborhoods can inform transport planning,
but detailed observations are often difficult to obtain. 
\emph{Travel surveys}
record trip purpose and transportation mode, yet collecting diaries
for an entire city is costly and time-consuming \cite{stopher2007}.
\emph{Movement traces} reveal regularities in human movement
\cite{gonzalez2008,song2010predictability,schneider2013}, but trip purpose
and mode often require inference \cite{shen2014}, and access to
individual trajectories raises privacy concerns \cite{demontjoye2013}.
Even when movement traces are available, their interpretation for urban
policy requires connecting observed flows to activity destinations,
neighborhoods, and transportation systems \cite{chang2021}.
These constraints motivate the use of population and infrastructure data
to investigate geographic differences in mobility where detailed travel
observations are limited.

Cities also contain substantial information that describes the conditions
for movement without directly recording it. Census statistics characterize
the population and its spatial distribution \cite{ibge2022}; building
locations and land-use data identify potential destinations
\cite{klinkhardt2021}; and street and transit networks describe how these
places are connected \cite{boeing2017,horl2021}. These sources are generally
aggregated or describe the city at a particular point in time. A census
tract may indicate how many people live in an area, for example, but does
not specify where they work or study, when they travel, or which mode
they use. Studies of urban form establish relationships between travel,
density, land-use mix, and accessibility
\cite{cervero1997,ewing2010,geurs2004}, but translating these relationships
into individual activity sequences requires an explicit model of travel
demand.

Synthetic-population methods provide a link between aggregate descriptions
and individual behavior by constructing records consistent with known
population totals \cite{beckman1996,ye2009,farooq2013}. Placing these
individuals within a geographic representation of the city allows them
to be assigned homes, workplaces, schools, and other destinations.
Activity-based models then represent travel as a consequence of
participation in daily activities \cite{bowman2001,arentze2000,bhat2004}.
Open-data pipelines have demonstrated the generation of detailed synthetic
travel demand \cite{horl2021}, and recent patterns-of-life simulation has
extended activity generation to synthetic populations across multiple
cities \cite{reia2026}. Probabilistic choice models connect these trips
to transportation alternatives through their relative attractiveness
\cite{mcfadden1974,benakiva1985}. Together, these approaches make it possible
to examine how the spatial organization of activities and transport access
shapes mobility. Building on them,
we ask how well their integration captures aggregate transit geography
and what it reveals about spatial differences in mode choice and
responses to improved transit access.

Here, we build on our patterns-of-life simulator \cite{reia2026} 
by combining its activity-generated trips with
road and bus-network travel times and probabilistic assignment of
walking, car, or public transit. Our contribution is the integration of
these components for spatial analysis of transportation choice and
controlled accessibility experiments. We evaluate the resulting
neighborhood-level transit patterns against an empirical activity proxy,
characterize the spatial distribution of mode shares, and examine how
those shares respond to reduced walking times to and from bus stops.
The aggregate comparison assesses broad spatial correspondence rather
than the accuracy of individual travel decisions \cite{drchal2016}.

We apply the framework to Niter\'oi, Brazil, where the distribution of
activities and bus infrastructure creates contrasting travel
opportunities across the municipality. The simulation identifies
walking-oriented central areas, car-oriented peripheral areas, and
areas with comparatively high transit use along bus corridors. A
counterfactual reduction in first- and last-mile travel times produces
larger transit-share gains in several peripheral areas. These results
illustrate how the framework can identify spatial differences in
mobility and explore where improved transit access may have the
greatest modeled effect.

\begin{figure}[t!]
  \centering
  \includegraphics[width=0.435\linewidth]{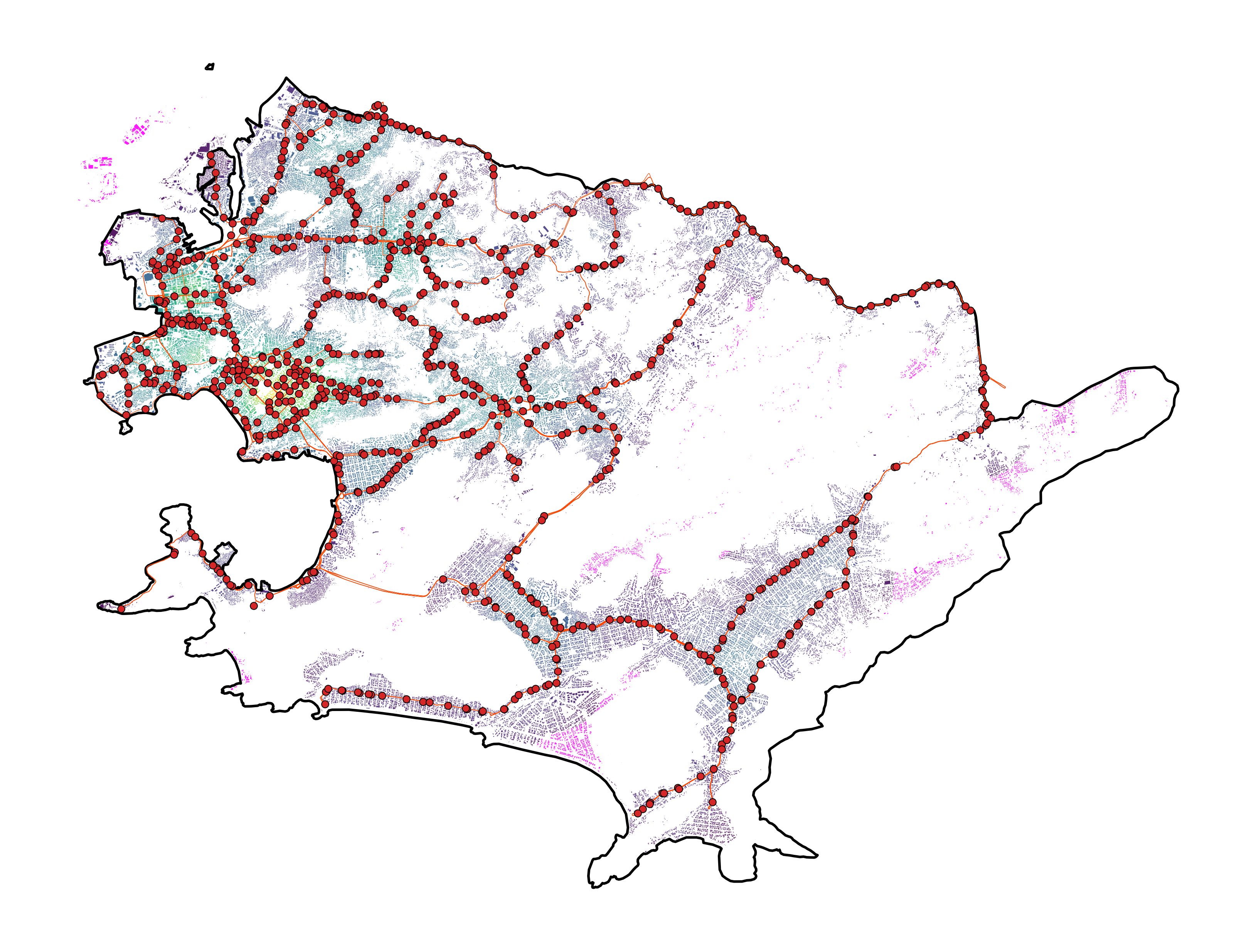}
  \includegraphics[width=0.545\linewidth]{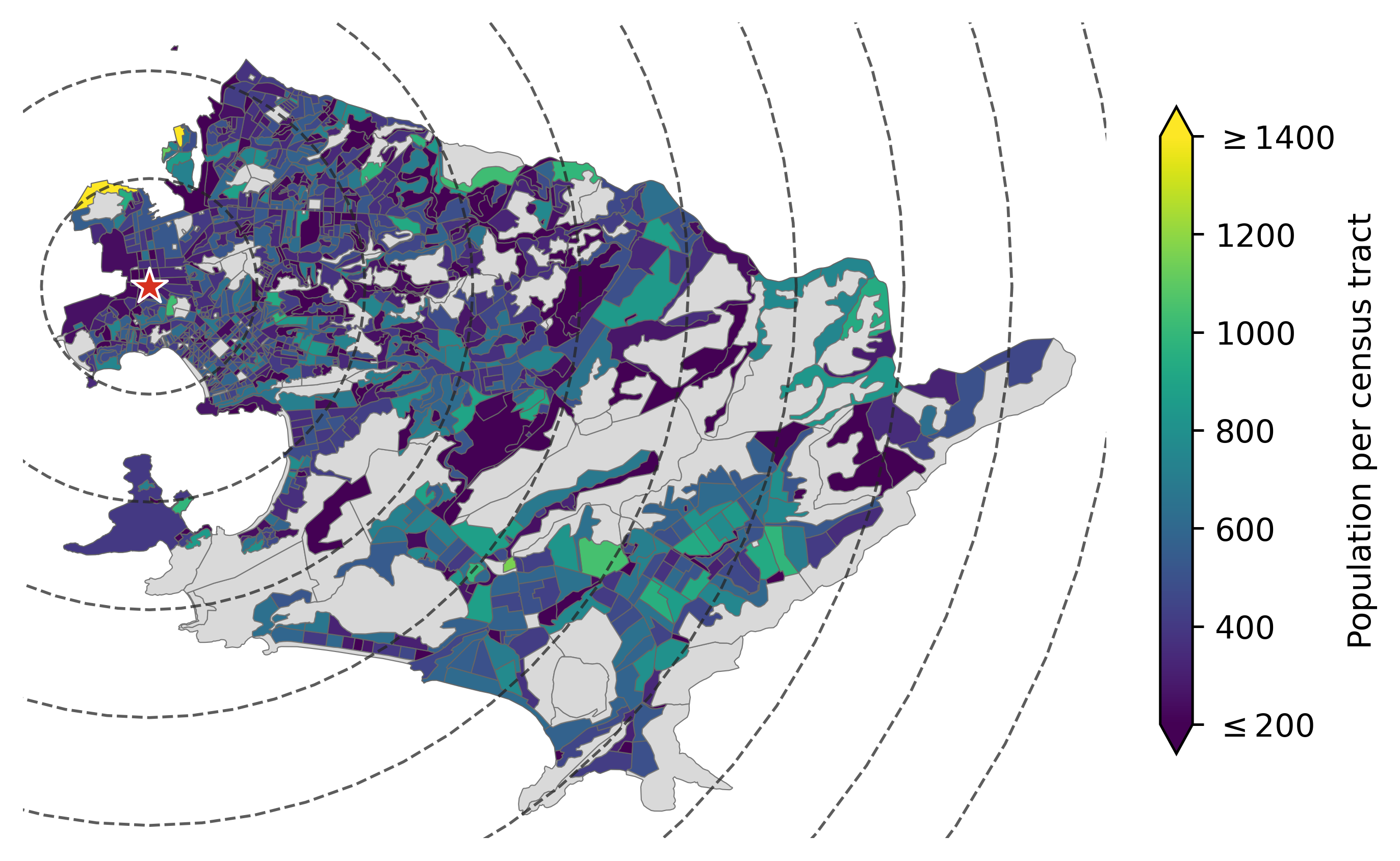}
    \caption{\textbf{Spatial heterogeneity of infrastructure, population and public-transit coverage of Niter\'oi.} The left panel maps the municipal bus network and stop locations. Orange lines represent bus routes, and red points represent bus stops. Building footprints are colored by the number of stops within a 10-minute walk (at \SI{5}{\kilo\metre\per\hour}). Colors progressing from purple and blue to green and yellow indicate increasing stop availability, while magenta identifies buildings with no stop within this distance. The right panel shows the synthetic resident population assigned to each census tract. Gray tracts contain no represented residents. The red star marks the centroid of the tract with the highest density of non-residential buildings, which is used as the city center. Dashed rings mark successive \SI{2}{\kilo\metre} distances from this point and provide a descriptive center--periphery coordinate.}
  \label{fig:study_area}
\end{figure}

\section{Methods}
\label{sec:data}

Our modeling workflow links population and urban infrastructure data to
individual activities, trips, and transportation choices. We first
describe the study area and the data used to construct the building
layer and synthetic population. We then explain how the activity
simulator generates trips and how road and bus-network travel times
enter the probabilistic assignment of walking, car, or public transit.
Together, these steps provide the baseline mobility representation used
to evaluate aggregate transit patterns and explore changes in transit
accessibility.

\subsection{Study area}

Niter\'oi is a coastal municipality in the metropolitan area of Rio de Janeiro, Brazil. 
Its irregular shoreline, protected areas, low-density settlements, and concentration of services along elongated urban corridors create sharp spatial differences in population density and transport access (\fig{fig:study_area}). 
According to the 2022 Brazilian Census, the municipality had 481,749 residents, of whom approximately 54.2\% were female and 23.8\% were aged 60 years or older \citep{ibge2022census}.
In the synthetic population used here, more than 81\% of residents live
within \SI{8}{\kilo\metre} of the city center
(\fig{fig:study_area}, Supplementary Fig.~S1).
Earlier local studies have identified strong neighborhood contrasts in trip generation and the potential for walking and public transport \citep{grieco2010trip,grieco2020built}. For regional context, the TUMI report presents a 2015 modal split for the Rio de Janeiro metropolitan region of 44.8\% public transport, 26.9\% walking, 24.3\% car travel, and 4.0\% cycling \citep{tumi2020niteroi}. These figures highlight the substantial role of public transport and walking in metropolitan mobility. In Niter\'oi, municipal planning has framed a shift toward sustainable mobility as a long-term policy objective, prioritizing public transport and active travel \citep{barandier2015backcasting,tumi2020niteroi}.

\subsection{Model framework and data layers}

\begin{figure}[t!]
  \centering
  \includegraphics[width=0.65\linewidth]{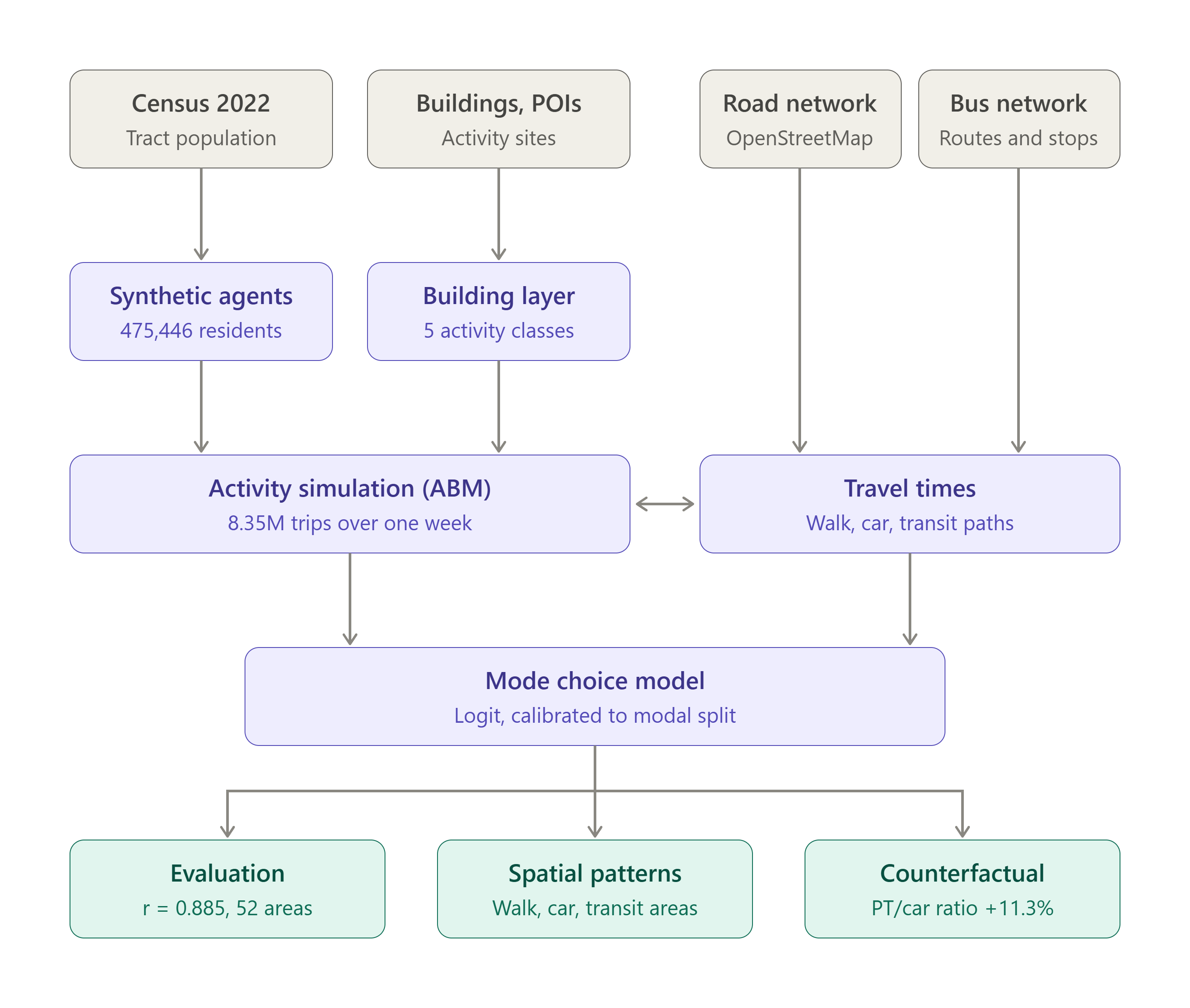}
    \caption{\textbf{Overview of the modeling framework.}
    Open data inputs (top) include census-tract population, building footprints and points of interest, the road network, and the bus network. Census data are used to construct a synthetic population of 475,446 agents, while building and facility data define a building layer with five activity classes. The agent-based activity simulation combines these inputs to generate 8.35 million trips over one week and is coupled with door-to-door travel times for walking, car, and public transit, derived from the road and bus networks. A multinomial logit model, calibrated to the regional modal split, assigns a transportation mode to each trip. The resulting mobility patterns support three analyses (bottom): evaluation against a public-transit activity proxy, identification of spatial patterns of transportation choice, and a counterfactual scenario.}
  \label{fig:framework}
\end{figure}

Our framework combines a representation of the population and its activity
locations with the transport networks connecting those locations.
Population and building data support the generation of activity schedules
and trips, while road and bus-network data provide the travel times used
to assign transportation modes. This workflow is summarized in
\fig{fig:framework}, with the principal inputs and derived datasets listed
in Table~\ref{tab:data}.

Census-tract boundaries and population composition were obtained from the 2022 Brazilian Population Census \citep{ibge2022census} and used to construct the synthetic population.
Building footprints were obtained from the Google--Microsoft--OpenStreetMap Open Buildings collection compiled by VIDA and distributed through Source Cooperative \citep{vida2024openbuildings}, while points of interest were obtained from the municipal Niter\'oi HUB SIGeo portal \citep{sigeoniteroi}. Together, these sources define the building layer, which identifies where agents live and perform activities.

The road network was obtained from OpenStreetMap \citep{openstreetmap} and used to calculate travel distances and estimate travel times for car and walking trips.
The bus network was assembled from municipal datasets available through SIGeo. Route geometries were obtained from the RotasMobNit layer \citep{sigeorotasmobnit}, while stop locations, the bus-line catalog, and stop--line associations were obtained from the Paradas de \^Onibus service \citep{sigeoparadas}. The road and bus networks are combined to estimate door-to-door travel times for each mode.

The synthetic population and building layer were supplied to the activity
simulation framework described in \citet{reia2026} to generate schedules
and trips. Each trip was then evaluated against the available transport
alternatives using their estimated travel times, and a probabilistic
model assigned its transportation mode. The following subsections describe
the construction of these inputs and the activity-generation and
mode-assignment procedures.

%


\begin{table}[htbp]
  \centering
  \small
  \caption{Data layers, their sources, and analytical roles.}
  \label{tab:data}
  \begin{tabular}{p{0.22\linewidth}p{0.25\linewidth}p{0.43\linewidth}}
    \toprule
    Layer & Source & Role \\
    \midrule

    Census data &
    IBGE 2022 Census \citep{ibge2022census} &
    Population composition, census-tract boundaries, and
    aggregation units. \\

    Building footprints &
    VIDA / Source Cooperative \citep{vida2024openbuildings} &
    Building geometries supporting 104,041 sampled residential
    and activity locations. \\

    Points of interest &
    Niter\'oi HUB SIGeo \citep{sigeoniteroi} &
    Municipal facility locations used to identify activities
    supported by buildings. \\

    Road network &
    OpenStreetMap \citep{openstreetmap} &
    Network used to calculate travel distances and estimate
    travel times for car and walking trips. \\

    Bus-route geometries &
    SIGeo: RotasMobNit \citep{sigeorotasmobnit} &
    Route geometries, directions, and line identifiers used
    to construct the bus network. \\

    Bus stops and line associations &
    SIGeo: Paradas de \^Onibus \citep{sigeoparadas} &
    Stop locations, line catalog, and stop--line associations
    supporting transfer-aware routing and travel-time estimation. \\

    Synthetic population &
    Simulation framework \citep{reia2026} &
    475,446 agents with demographic attributes, activity roles,
    assigned locations, needs, and schedules. \\

    Simulated events and derived trips &
    Simulation framework \citep{reia2026} &
    Activity states and building changes yielding 8,350,189
    trip records over seven days, including times, origins,
    and destinations. \\

    \bottomrule
  \end{tabular}
\end{table}

\subsubsection{Building infrastructure}
\label{subsec:building_infrastructure}

The building layer is needed by the ABM activity generation simulation framework we use here \cite{reia2026} to define the locations where agents reside and perform activities. 
Building footprints were obtained from the Brazilian subset of the Google--Microsoft--OpenStreetMap Open Buildings collection, prepared by VIDA and distributed through Source Cooperative \citep{vida2024openbuildings}. 
Because the footprint dataset provides limited information on building use, we supplemented it with 13 municipal facility layers downloaded as GeoJSON files from Niter\'oi HUB SIGeo \citep{sigeoniteroi}. These layers identify facilities such as police stations, companies, schools, universities, pharmacies, hospitals, tourist attractions, gas stations, and restaurants.

Each facility point was matched to the building footprint containing it, including points on the polygon boundary. Points outside all footprints were matched to the nearest building. These matches transferred facility classifications to the building layer. Buildings without an identified nonresidential facility were classified as residential.
The resulting layer comprises the five activity classes used in the simulation: residential buildings, workplaces, food-service establishments, educational facilities, and recreational facilities.

\subsubsection{Synthetic population}

We constructed a synthetic population for Niter\'oi using census-tract data from the 2022 Brazilian Population Census \citep{ibge2022census}. 
The synthetic population is a key input to the ABM simulation framework we use here \cite{reia2026}.
We created one agent for each resident counted in 11 age groups: 0--4, 5--9, 10--14, 15--19, 20--24, 25--29, 30--39, 40--49, 50--59, 60--69, and 70 years or older. 
Ages in the open-ended group were sampled between 70 and 80 years. Sex was assigned independently of age and tract using the municipal proportions of 45.8\% male and 54.2\% female \citep{ibge2022census}.
Each agent was assigned a home in its census tract. Building footprints were projected to WGS~84/UTM zone 23S (EPSG:32723) to calculate their areas in square metres, which was used as a proxy for residential capacity and allows multiple agents to share a building. 
Agents in tracts without an eligible building were excluded, yielding a final population of 475,446 agents. 

Agents were assigned one of three roles---worker, student, or
homemaker---based on their age and gender, while matching the demographic
composition and employment rate of their census tract
\citep{ibge2022census}.
Agents without a workplace were classified as students if they were
17 years old or younger, and as homemakers otherwise.
Here, ``homemaker'' is a model label for adults without an assigned
workplace; it does not indicate their actual occupation or employment
status. 
Workers and students were then assigned regular destinations from the building layer. 
Each worker received a randomly selected workplace within \SI{15}{\kilo\metre} of home while each student was assigned to the nearest school. 

Each agent was assigned a fixed set of 12 restaurants, 12 recreational facilities, and 26 possible errand destinations available throughout the simulation.
Limiting these sets was motivated by evidence that people regularly
return to a relatively small set of familiar places
\citep{alessandretti2018conserved}. 
Whenever a discretionary destination was needed, it was selected from
the corresponding agent-specific list using a Zipf distribution with
exponent one: the probability of selecting a destination of rank $r$
was proportional to $1/r$ \citep{zipf1949}. Thus, the first-ranked
destination was twice as likely to be selected as the second-ranked
destination and three times as likely as the third-ranked destination.
Rank represents modeled preference, producing unequal expected visitation
frequencies consistent with the tendency to concentrate visits on a few
preferred locations \citep{song2010scaling}.


\subsection{Activity schedules}

\begin{figure}
    \centering
    \includegraphics[width=1\linewidth]{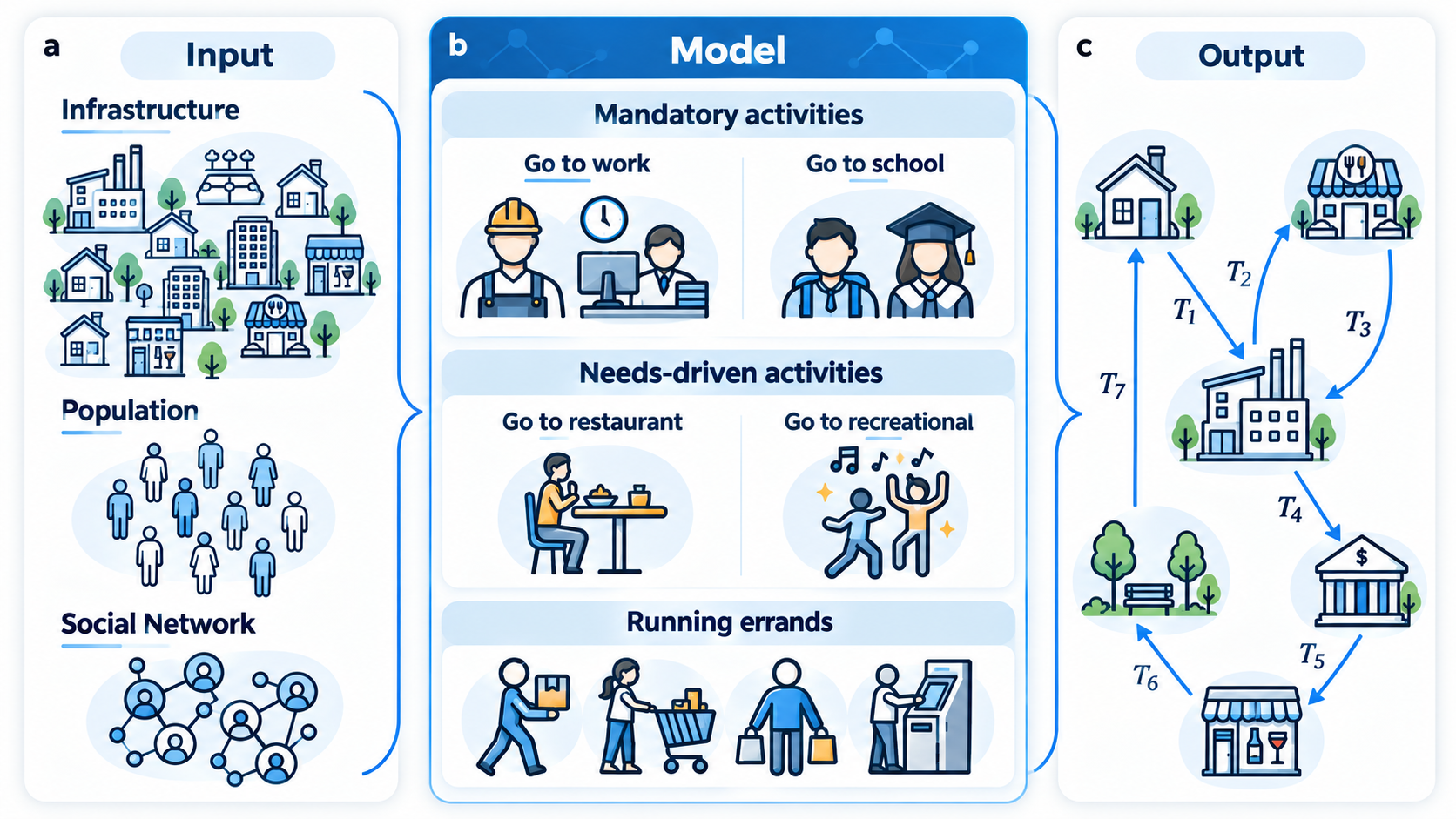}
    \caption{\textbf{Agent-based model used to generate activity schedules.}
    The model combines information on the synthetic population and urban infrastructure to generate individual sequences of activities and trips. Mandatory activities are associated with fixed destinations, such as workplaces and schools, while needs-driven activities, including restaurant visits, recreation, and errands, are assigned using simple behavioral heuristics and agent-specific sets of possible destinations. The resulting sequence of visited locations defines the daily activity schedule, with transitions between consecutive activities generating trips between buildings. Adapted from \cite{reia2026}.}
    \label{fig:ABM}
\end{figure}

Activity schedules were generated using the agent-based simulation framework introduced in \citet{reia2026}. 
Agent based simulations have been used to describe individuals' behavior in terms of needs ranging from basic survival to social belonging and self-fulfillment, determining agent's routine and schedules in urban simulations \cite{zufle2023urban}. 
As illustrated in Fig.~\ref{fig:ABM}, the model combines information about the synthetic population and urban infrastructure to generate sequences of activities for each agent. Activities are divided into two broad classes. Mandatory activities are determined by the agent's role: workers travel to their assigned workplace and students to their assigned school. Needs-driven activities, including eating at restaurants, recreation, and errands, can occur at other destinations available to the agent.

Needs-driven activities use the agent-specific destination sets and
selection rule described above. These behavioral heuristics represent
recurring destination preferences without attempting to reproduce the
exact decisions of individual residents.

The resulting schedule consists of an ordered sequence of activities performed at specific buildings and times. Whenever two consecutive activities occur at different buildings, the transition between them defines a trip, including its origin, destination, departure time, and arrival time. The simulation therefore transforms static population and infrastructure data into a dynamic representation of individual activity and mobility patterns. The number of trips emerges from both scheduled commitments, such as work and school, and discretionary activities generated by the accumulation of food, social, and errand needs. When these needs exceed their respective thresholds, they trigger activities at suitable destinations and consequently generate additional trips. The parameters controlling this process were calibrated against the mobility index reported in the Sustainable Urban Mobility Plan of Niter\'oi (PMUS), which estimates an average of \(2.49\) trips per inhabitant per day across all transportation modes in 2018 \cite[p.~40]{PMUSNiteroi2020}. Over the seven-day simulation, the model generated \(8{,}350{,}189\) trips for \(475{,}446\) agents, corresponding to an average of \(2.51\) trips per agent per day. 

\subsection{Probabilistic mode assignment}
\label{sec:probModel}

Transportation modes were assigned using a multinomial logit model,
a standard approach to discrete choice and travel-demand modeling
\citep{mcfadden1974conditional,benakiva1985discrete,train2009discrete}.
This formulation allows alternatives with shorter travel times to
be more likely without assuming that every traveler selects the
fastest mode.
For trip $i$ and available mode $m$, the systematic utility and choice
probability are given by
\begin{align}
V_{im} &= \alpha_m + \beta_t t_{im}, \label{eq:utility}\\
P_{im} &= \frac{\exp(V_{im})}
{\sum_{k\in\mathcal{A}_i}\exp(V_{ik})}, \label{eq:mnl}
\end{align}
where $\mathcal{A}_i$ denotes the choice set available for trip $i$,
comprising walking, car, and public transit (PT). The travel time
$t_{im}$ varies across trips and modes, whereas $\beta_t$ is a common
travel-time coefficient. Its negative value represents the disutility of
travel time: all else being equal, alternatives with longer travel times
are less likely to be selected. Using a common coefficient assumes that
an additional minute has the same marginal effect on utility for all
three modes. With $\beta_t=-0.10$, 
one additional minute reduces utility
by $0.10$ and multiplies the odds of choosing that mode by
$\exp(-0.10)\simeq0.905$, corresponding to an approximately 9.5\%
reduction in its choice odds, holding the other alternatives unchanged.

The alternative-specific constants $\alpha_m$ capture differences in the
average attractiveness of each mode that are not explained by modeled
travel time. At an aggregate level, they therefore absorb omitted factors
such as monetary cost, comfort, convenience, reliability, flexibility,
parking conditions, waiting conditions, and prevailing travel habits.
Because only differences in utility are identifiable, the car constant is
fixed at zero and serves as the reference. Consequently,
$\alpha_{\mathrm{walk}}$ and $\alpha_{\mathrm{PT}}$ represent the baseline
utilities of walking and public transit relative to car after accounting
for differences in travel time.

The mode-assignment parameters were calibrated against the aggregate modal
split reported by TUMI for the Metropolitan Region of Rio de Janeiro in
2015 \cite{tumi2020niteroi}: 26.9\% walking, 24.3\% car, 44.8\% public
transit, and 4.0\% cycling. Because cycling is not represented as a separate
mode in the model, its share was combined with walking for calibration.
The resulting targets were therefore 30.9\% for walking, 24.3\% for car,
and 44.8\% for public transit. These citywide shares are calibration targets,
not validation outcomes; the model is evaluated below on how public-transit
activity is distributed spatially across Niter\'oi.

Walking and driving times are approximated from the trip distance $d_i$
using mode-specific constant speeds:
\begin{equation}
t_{im}=\frac{d_i}{v_m},
\qquad
m\in\{\mathrm{walk},\mathrm{car}\},
\label{eq:simple_times}
\end{equation}
with distances and speeds converted to obtain travel times in minutes.
The assumed speeds are
$v_{\mathrm{walk}}=\SI{5}{\kilo\metre\per\hour}$ and
$v_{\mathrm{car}}=\SI{30}{\kilo\metre\per\hour}$.
Public-transit travel times explicitly account for access, in-vehicle
travel, transfers, and egress. Each origin and destination building is
connected to its nearest bus stop. Transit paths are calculated on a
directed line-state graph whose nodes retain both the traveler's physical
position and the bus line currently occupied. In-vehicle edges assume a
speed of \SI{20}{\kilo\metre\per\hour}, while each change of line incurs
a transfer penalty of \SI{10}{\minute}.
The resulting door-to-door public-transit travel time is
\begin{equation}
t_{i,\mathrm{PT}}
=
t_{i,\mathrm{access}}
+t_{i,\mathrm{ride}}
+\tau n_{i,\mathrm{transfer}}
+t_{i,\mathrm{egress}},
\label{eq:transit_time}
\end{equation}
where $n_{i,\mathrm{transfer}}$ is the number of line changes and
$\tau=\SI{10}{\minute}$ is the penalty associated with each transfer.
Public transit is excluded from $\mathcal{A}_i$ when no directed path
connects the selected origin and destination stops.

The calibrated specification uses $\beta_t=-0.10$,
$\alpha_{\mathrm{walk}}=6.1052$, $\alpha_{\mathrm{car}}=0$, and
$\alpha_{\mathrm{PT}}=6.3541$. Calibration matches the citywide mean
choice probabilities to the target shares. Sampling modes from these
probabilities yields realized shares of 30.8989\% for walking,
24.3036\% for car, and 44.7961\% for public transit. These differ from
the targets by $-0.0011$, $+0.0036$, and $-0.0039$ percentage points,
respectively. This close agreement is a calibration check rather than
an independent validation result.

\section{Results}

In the following we examine how the simulated activities translate into spatial patterns
of travel demand and transportation choice in Niter\'oi. We first
characterize activity patterns and the associated redistribution of
population and trip distances across the city. We then compare aggregate
public-transit activity with the empirical proxy and describe geographic
differences in the shares of walking, car, and public transit. Finally,
we examine how reducing transit access and egress times changes these
shares and where the modeled response is strongest.

\subsection{Activity patterns and trip chains}

\begin{figure}
    \centering
    \includegraphics[width=1\linewidth]{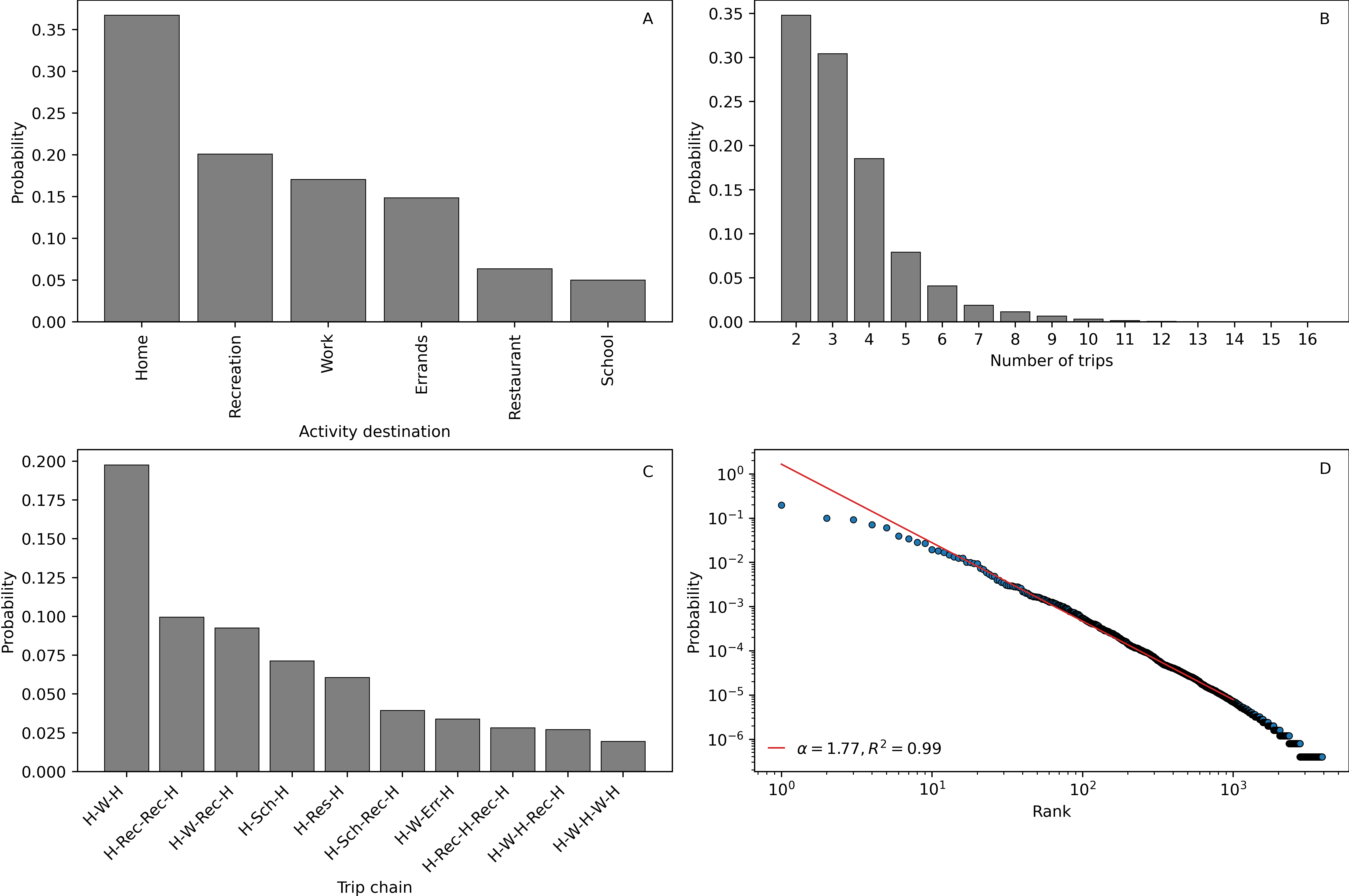}
    \caption{\textbf{Activity patterns generated by the agent-based model.}
    All panels summarize the simulated schedules of the 475,446 synthetic agents over the seven-day simulation (8.35 million trips). 
    (A) Share of all simulated trips by destination category. Home is the most frequent destination (${\sim}37\%$), followed by recreation (${\sim}20\%$), work (${\sim}17\%$), errands (${\sim}15\%$), restaurants (${\sim}6\%$), and school (${\sim}5\%$). Home dominates because every chain ends with a return home.
    (B) Distribution of chain length, measured as the number of trips per chain. Chains are short: about 35\% contain two trips (a single outing and the return) and 30\% contain three. Chains with five or more trips account for less than 20\% of the total, and the longest contains 16 trips.
    (C) The ten most probable chains. H--W--H alone accounts for about 20\% of all chains, and the ten chains together account for about 67\%. H = Home, W = Work, Sch = School, Rec = Recreation, Res = Restaurant, Err = Errands.
    (D) Rank--probability plot (log--log axes) of all distinct chains, where rank 1 is the most probable chain. The red line is a linear fit in log--log space, with slope $\alpha = 1.77$ and $R^2 = 0.99$, used here as a descriptive summary of the strong rank decay rather than as a formal test of power-law behavior.}
    \label{fig:activity_patterns}
\end{figure}

The agent-based model generates the daily activity schedules of each agent in Niter\'oi using six activity types: Home, Work, School, Restaurant, Recreation, and Errands. Since agents always start and end their daily activity schedules at home \cite{primerano2008defining}, home naturally plays a central role in the resulting mobility patterns (Fig.~\ref{fig:activity_patterns}), consistent with previous findings \cite{reia2025function}.
The distribution of trips across activity types, particularly Work and School, is strongly influenced by agent typology \cite{reia2026}. Workers represent approximately 47.73\% of the synthetic population, resulting in about 20\% of all trips being associated with workplace activities. In contrast, students account for approximately 17.6\% of the population, and trips to school represent only about 5\% of the total.

The relatively large number of trips associated with Recreation and Errands is partly explained by the fraction of homemakers in the population, which is approximately 34.7\%. Homemakers are agents without mandatory activities such as Work or School and represent, for example, retired individuals, stay-at-home parents, and unemployed people. The number of activities that an agent can perform each day is constrained by the time available in their schedule. Because homemakers do not have mandatory activities occupying fixed portions of the day, they have greater flexibility to accommodate needs-driven activities, including Recreation and Errands \cite{reia2026}.

The importance of work in daily mobility is also reflected in the most common trip chains (Fig.~\ref{fig:activity_patterns}C). A trip chain is defined as the ordered sequence of activity locations visited by an agent during a day, beginning and ending at home \cite{primerano2008defining}. The most common chain is H--W--H, representing a direct trip from home to work followed by the return home, and accounts for almost 20\% of all daily chains. Other work-centered patterns, such as H--W--Rec--H and H--W--Err--H, also appear among the ten most frequent chains, highlighting the influence of worker agents on the overall structure of daily mobility.

The effect of agent typology is also evident in other recurrent chains. For example, H--Sch--H captures the typical daily mobility of students and is among the most common patterns. At the same time, the relatively large fraction of homemakers contributes to the prevalence of chains containing no mandatory Work or School activities, such as H--Rec--Rec--H and H--Res--H. These patterns reflect the greater scheduling flexibility of agents without fixed mandatory activities, who can devote a larger fraction of their daily schedules to discretionary destinations such as Recreation, Restaurants, and Errands.
Despite the recurrence of a small number of dominant trip chains, the full distribution is highly heterogeneous. The probability of observing a given chain decreases steeply with rank (Fig.~\ref{fig:activity_patterns}D), producing a long-tailed rank--frequency pattern. A linear fit in log--log space gives a slope of $\alpha = 1.77$ with $R^2 = 0.99$, which we use only as a descriptive summary. A relatively small number of common mobility patterns therefore account for a substantial fraction of daily schedules, while a long tail of less frequent chains captures the diversity of individual activity sequences across the population.

\subsection{Spatial representation of mobility needs}

The activity patterns and trip chains described above establish the
composition of simulated travel demand. We now examine where these
activities draw agents during the day and how trip distances vary across
Niter\'oi, providing the geographic context for the transportation choices
examined next.

\begin{figure}
    \centering
    \includegraphics[width=0.85\linewidth]{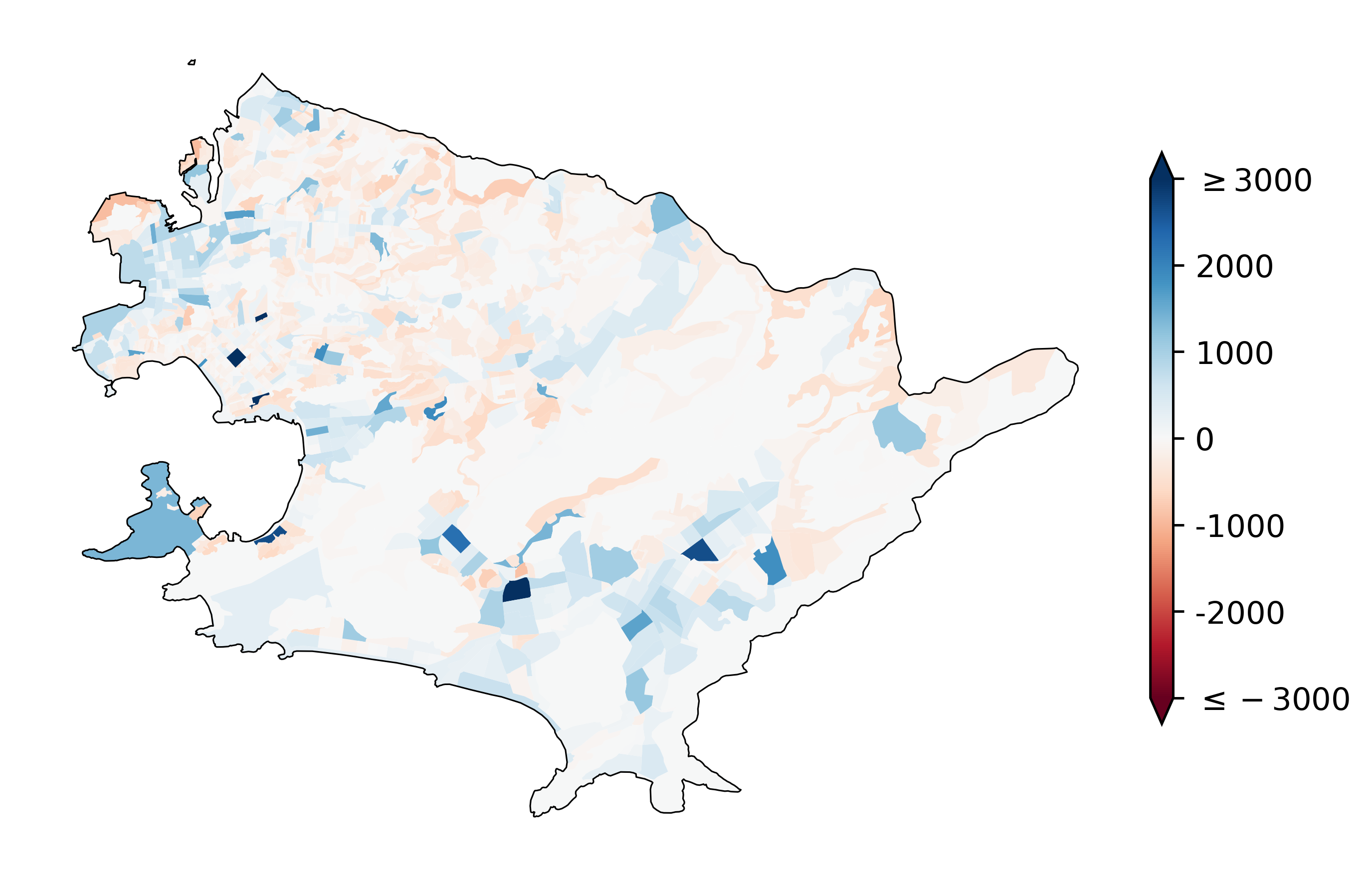}
    \caption{\textbf{Daily activities redistribute population toward activity centers.} The map shows the change in the number of agents present in each census tract between midnight and noon on a week day. 
    For each tract, the population shift is calculated as the number of agents present at noon minus the number present at midnight. Blue tracts gain population during the day, whereas red tracts lose population; lighter colors indicate little or no net change.}
    \label{fig:day_night_population_shift}
\end{figure}

The difference between the simulated populations present at noon and
midnight shows substantial daytime gains along the western, southern,
and southeastern portions of the shoreline
(Fig.~\ref{fig:day_night_population_shift}). These areas contain commercial
activities, services, employment opportunities, and other destinations
that attract agents away from their residential locations. Several
predominantly residential areas instead lose population during the day
as agents travel to their activity destinations.

\begin{figure}
    \centering
    \includegraphics[width=0.85\linewidth]{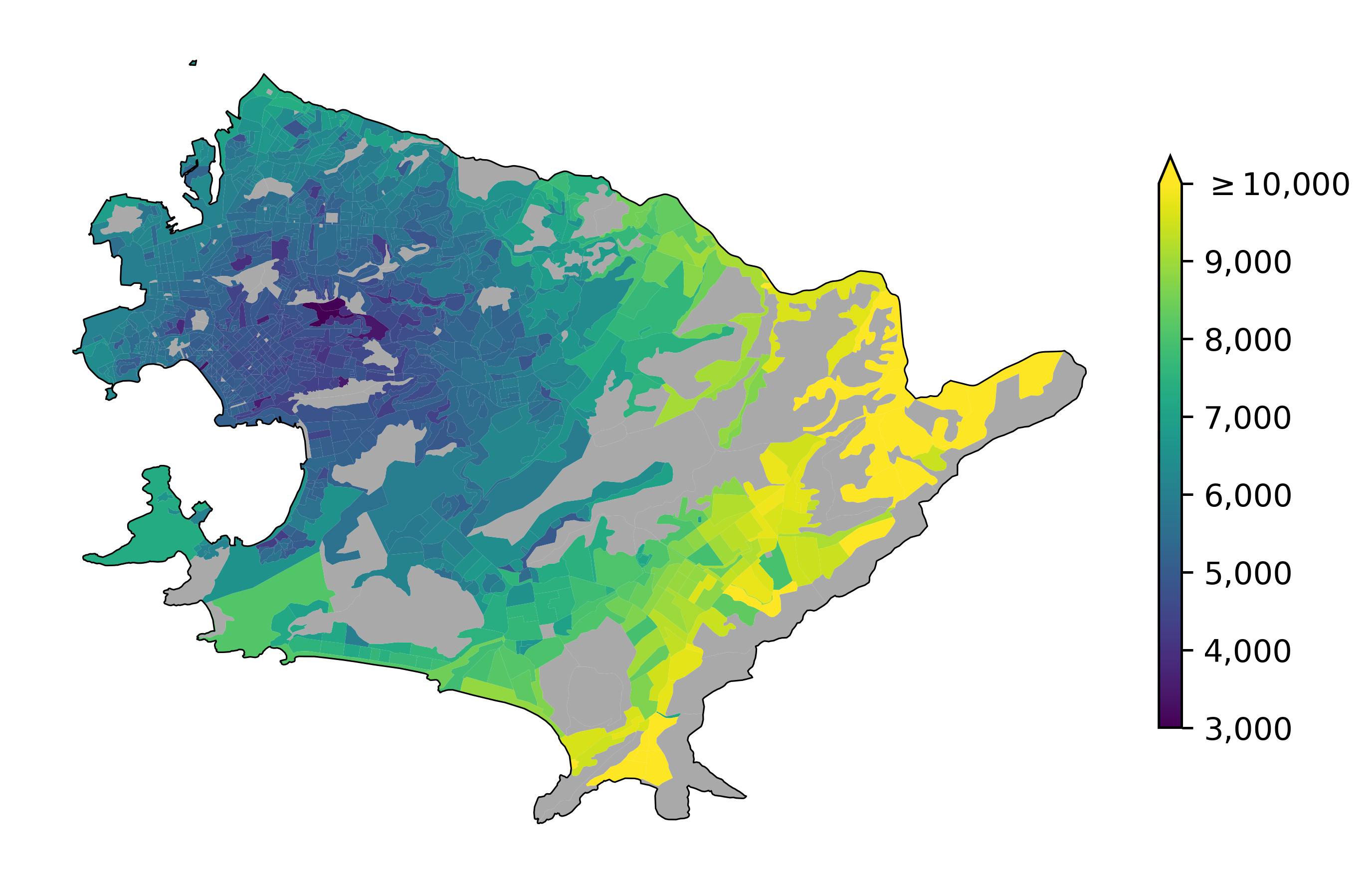}
    \caption{\textbf{Trip distances are spatially heterogeneous.} 
    The map shows the mean distance (in meters) of all simulated trips originating in each census tract. 
    Darker purple and blue colors indicate shorter mean distances, while green and yellow colors indicate longer distances. 
    Gray tracts contain no originating trips and are therefore treated as missing data.}
    \label{fig:spatial_heterogeneity_travel_distances}
\end{figure}

Beyond the redistribution of population, the trips connecting activity
locations reveal differences in travel distance across the city
(Fig.~\ref{fig:spatial_heterogeneity_travel_distances}).
Trips originating in census tracts located near the city center and the more urbanized western portion of Niter\'oi tend to be considerably shorter, with mean travel distances of approximately 3--6~km. These areas concentrate employment, services, and other activity opportunities, allowing agents to reach many destinations over relatively short distances. In contrast, mean trip distances generally increase toward the more peripheral eastern and northeastern portions of the municipality, where trips originating from some census tracts reach average distances of approximately 10--12~km.

\subsection{Transportation mode assignment}

The preceding analysis establishes where simulated activities draw agents
and how far they travel to reach them. We now examine how these trips
are distributed among walking, car, and public transit, given the
transport networks and travel times associated with their origins and
destinations. We first assess agreement with the neighborhood-level
transit-activity proxy, then examine how assigned mode shares vary with
distance from the city center and across census tracts.

Mode assignments follow the multinomial logit model described in
Eqs.~\ref{eq:utility} and \ref{eq:mnl}. Because detailed trip-level travel
data are not available for Niter\'oi, we assess the resulting patterns at
an aggregate scale by comparing simulated public-transit activity with a
neighborhood-level empirical proxy (Fig.~S2).

The simulated public-transit activity captures the broad spatial distribution represented by the empirical proxy across Niter\'oi. Both the empirical proxy and the simulation show that transit activity is concentrated in a relatively small number of neighborhoods in the western and central portions of the municipality, while most neighborhoods account for substantially smaller shares of citywide activity (Fig.~S2). 
This association is quantified in Fig.~S3, where the simulated share of public-transit activity in each of the 52 neighborhoods is compared with the corresponding image-derived activity share. The two distributions exhibit a strong linear association, with a Pearson correlation of $r=0.885$ ($p=3.39\times10^{-18}$). Their relative ordering across neighborhoods is similarly consistent, with a Spearman rank correlation of $\rho=0.873$ ($p=3.05\times10^{-17}$). This comparison supports the aggregate spatial distribution of simulated public-transit activity; it does not independently evaluate walking or car shares. 

\begin{figure}
    \centering
    \includegraphics[width=0.75\linewidth]{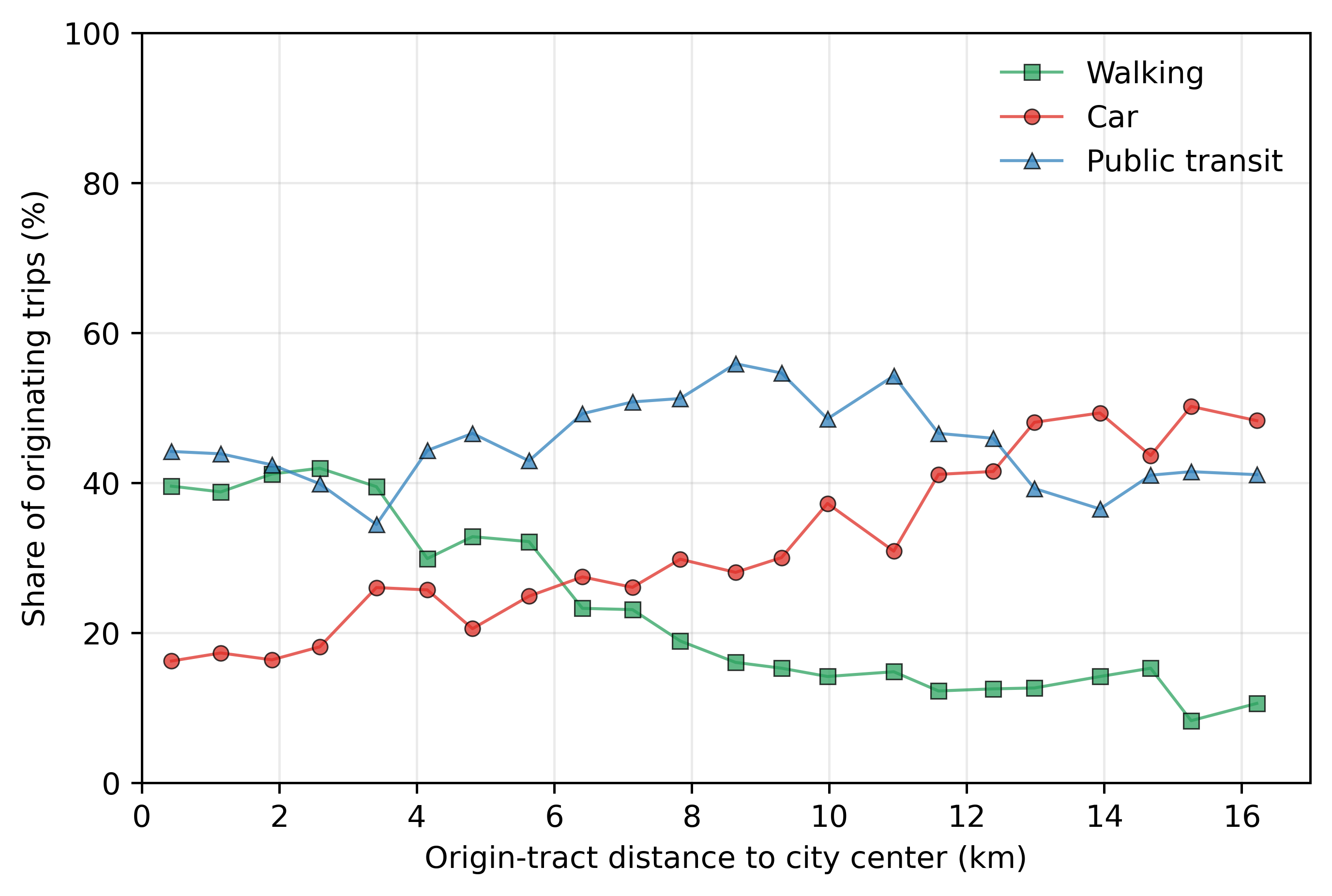}
    \caption{\textbf{Travel-mode choice varies with distance from Niter\'oi's city center.} 
    Simulated shares of trips assigned to walking, public transit, and car as a function of the distance between each trip's origin tract and the city center. 
    The city center is defined as the centroid of the census tract with the highest density of non-residential buildings. Census tracts are grouped into \(750\,\mathrm{m}\) distance bands, and the tract-level mode shares within each band are weighted by the number of trip ends represented in each tract.}
    \label{fig:share_transpode_city_center}
\end{figure}

Walking trips are predominantly short, with an average distance of approximately 2.51~km. 
Public-transit trips cover substantially longer distances, averaging 6.90~km, while car trips are the longest on average, reaching 9.83~km (Fig.~S4). 
These differences are also reflected in how transportation-mode choice varies with distance from Niter\'oi's city center (Fig.~\ref{fig:share_transpode_city_center}). 
Walking accounts for a large share of trips originating near the city center, where average trip distances are shorter. 
However, its share declines rapidly with distance, becoming the least-used mode in tracts farther from the city center. 
Public transit becomes the dominant mode across most intermediate distance bands, particularly between approximately \(4\) and \(12\,\mathrm{km}\). 
In contrast, car use generally increases toward the urban periphery and overtakes public transit beyond approximately \(13\,\mathrm{km}\). 

\begin{figure}
    \centering
    \includegraphics[width=1\linewidth]{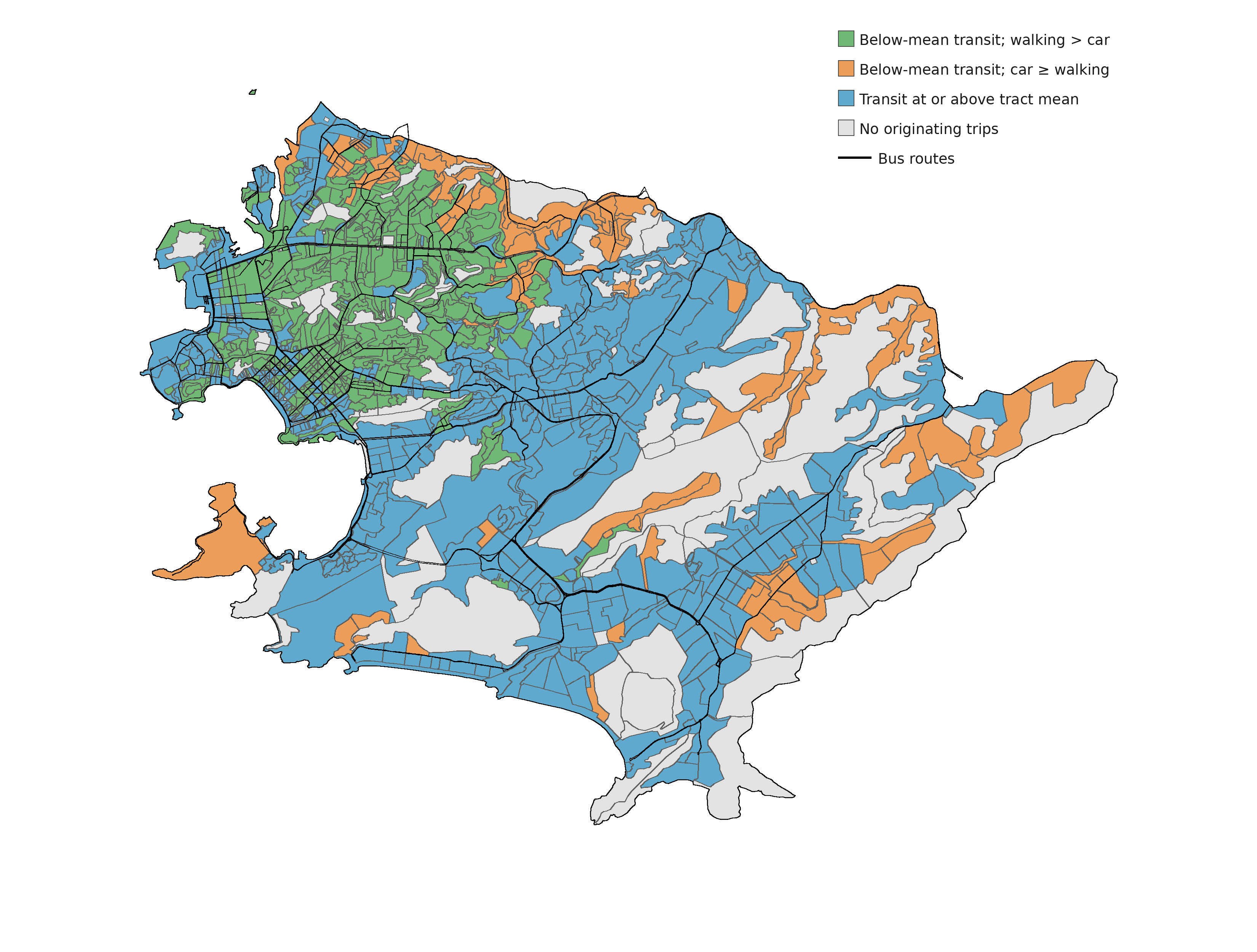}
    \caption{\textbf{Spatial patterns of transportation choice in Niter\'oi.}
    Walking is more prevalent in compact, service-rich areas, car use becomes more prominent in peripheral and lower-density tracts, and public transit remains important along corridors that connect both central and more distant neighborhoods. 
    These patterns show that transportation choice is shaped not only by distance from the city center, but also by the spatial organization of activities and transport accessibility.}
    \label{fig:spatial_patterns_transmode}
\end{figure}

To better capture the spatial patterns of transportation choice, we show in Fig.~\ref{fig:spatial_patterns_transmode} the map of Niter\'oi classified according to the dominant modal pattern of each census tract. The classification uses the citywide average public-transit (44.8\%, see Section \ref{sec:probModel}) as a reference. 
Tracts where public-transit use is equal to or above this average are shown in blue and identify areas where transit plays a comparatively important role in daily mobility. 
Among tracts with below-average public-transit use, green areas are those where walking is more prevalent than car travel, while orange areas are those where car use is more prevalent than walking. These categories distinguish whether walking or car travel is the stronger non-transit mode in each tract.

The resulting spatial structure reveals three broad mobility regimes across the city, which are also captured when examining the spatial distribution of each transportation mode individually (Fig.~S5). 
Walking-oriented tracts are concentrated mainly in the more compact and service-rich western portion of Niter\'oi, where activities are located closer together and shorter trips are more common. 
Car-oriented tracts appear more frequently in peripheral and lower-density areas, where agents generally need to travel longer distances to reach their destinations. 
Public-transit-oriented tracts extend across both central and more distant parts of the municipality, often following the structure of the bus network and connecting areas that are not necessarily close to the city center.


\subsection{Assessment of a transit-access scenario}

The spatial differences in transportation choice described above raise
the question of where improved transit access could make public transit
more competitive with car travel. The car-oriented tracts in
Fig.~\ref{fig:spatial_patterns_transmode} are of particular interest:
car use exceeds walking, while public-transit share remains below the
citywide reference. We examine the modeled response to improved access
through a counterfactual reduction in the walking times required to
reach and leave bus stops.

To motivate this scenario, we first examine how transit travel times
and their access and egress components vary across the city. This
identifies where reaching the bus network contributes substantially to
the travel burden of simulated transit trips and provides a basis for
interpreting the spatial response to the experiment.

\begin{figure}
    \centering
    \includegraphics[width=1\linewidth]{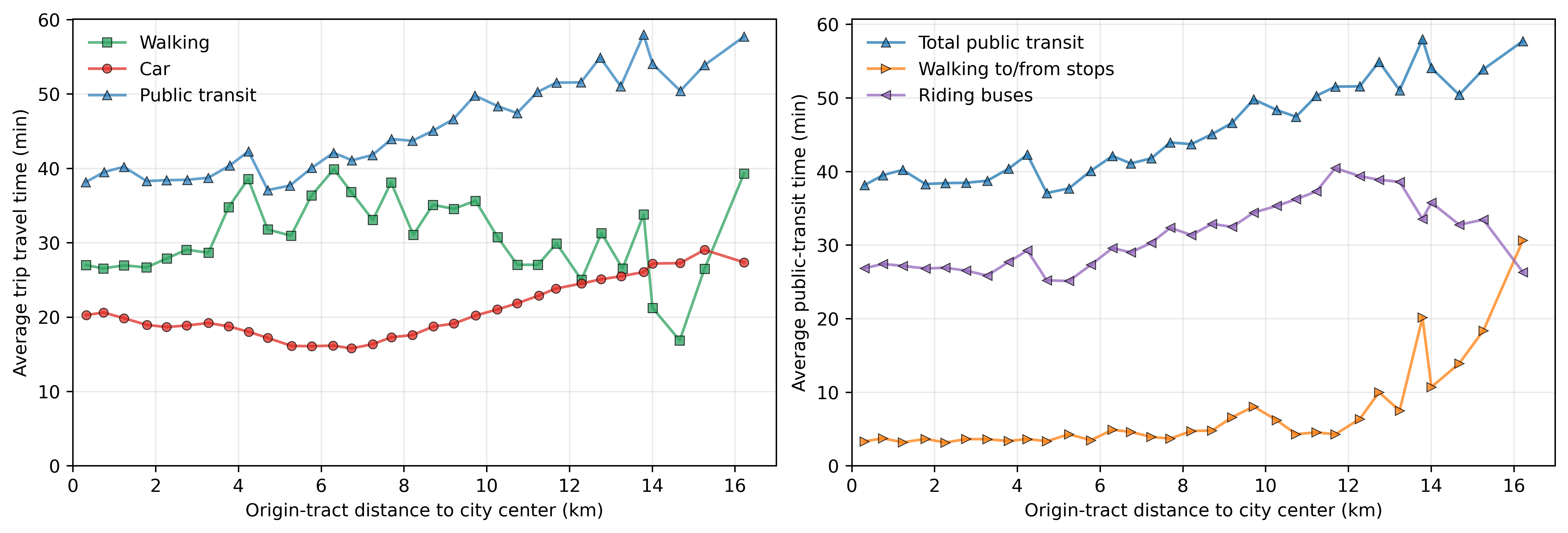}
    \caption{\textbf{Public transit and car travel times generally increase with distance from city center.} Trips are grouped into \SI{750}{\metre} intervals according to the distance between their origin census tract and the city center, defined as the centroid of the tract with the highest density of non-residential buildings. 
    Points show averages of the tract-level travel times within each interval. 
    The left panel compares trips according to their assigned transportation mode. 
     The right panel decomposes the travel time of trips assigned to public transit. Total public-transit time includes walking from the origin to a stop, riding buses, transfer penalties, and walking from the final stop to the destination.}
    \label{fig:travel_times_by_mode}
\end{figure}

Figure~\ref{fig:travel_times_by_mode} shows how the average travel time of assigned trips varies with the distance between their origin census tract and the city center. Car trips have the shortest average travel times across most distance bands, whereas trips assigned to public transit generally have the longest travel times and become progressively slower toward the urban periphery. Walking times are more variable but fluctuate around a mean of approximately 30~min.

As shown in Eq.~\ref{eq:transit_time}, total public-transit travel time can be decomposed into two broad components: access and egress time, associated with walking to the initial bus stop and from the final stop to the destination, and the time spent within the transit system, including bus riding and transfer penalties. This decomposition reveals an important spatial pattern. For trips originating within approximately 10~km of the city center, access and egress times remain relatively small, generally around a few minutes. Beyond this distance, however, they increase sharply, reaching approximately 30~min in the outermost distance band. In contrast, the time spent riding buses varies more gradually with distance. These results suggest that, in peripheral areas, a substantial portion of the total public-transit travel burden arises from reaching and leaving the bus network rather than from in-vehicle travel alone.

Motivated by this pattern, we design a simple counterfactual scenario to assess how improved access to public transit could affect transportation choice. Specifically, we reduce the access and egress components of public-transit travel time by 30\%, while leaving bus riding time, transfer penalties, trip destinations, and the travel times of the other transportation modes unchanged. This scenario does not represent a specific infrastructure project, but rather a reduction in first- and last-mile impedance that could approximate improvements such as better pedestrian connections, feeder services, stop accessibility, or other measures that facilitate access to the bus network.

\begin{figure}
    \centering
    \includegraphics[width=0.445\linewidth]{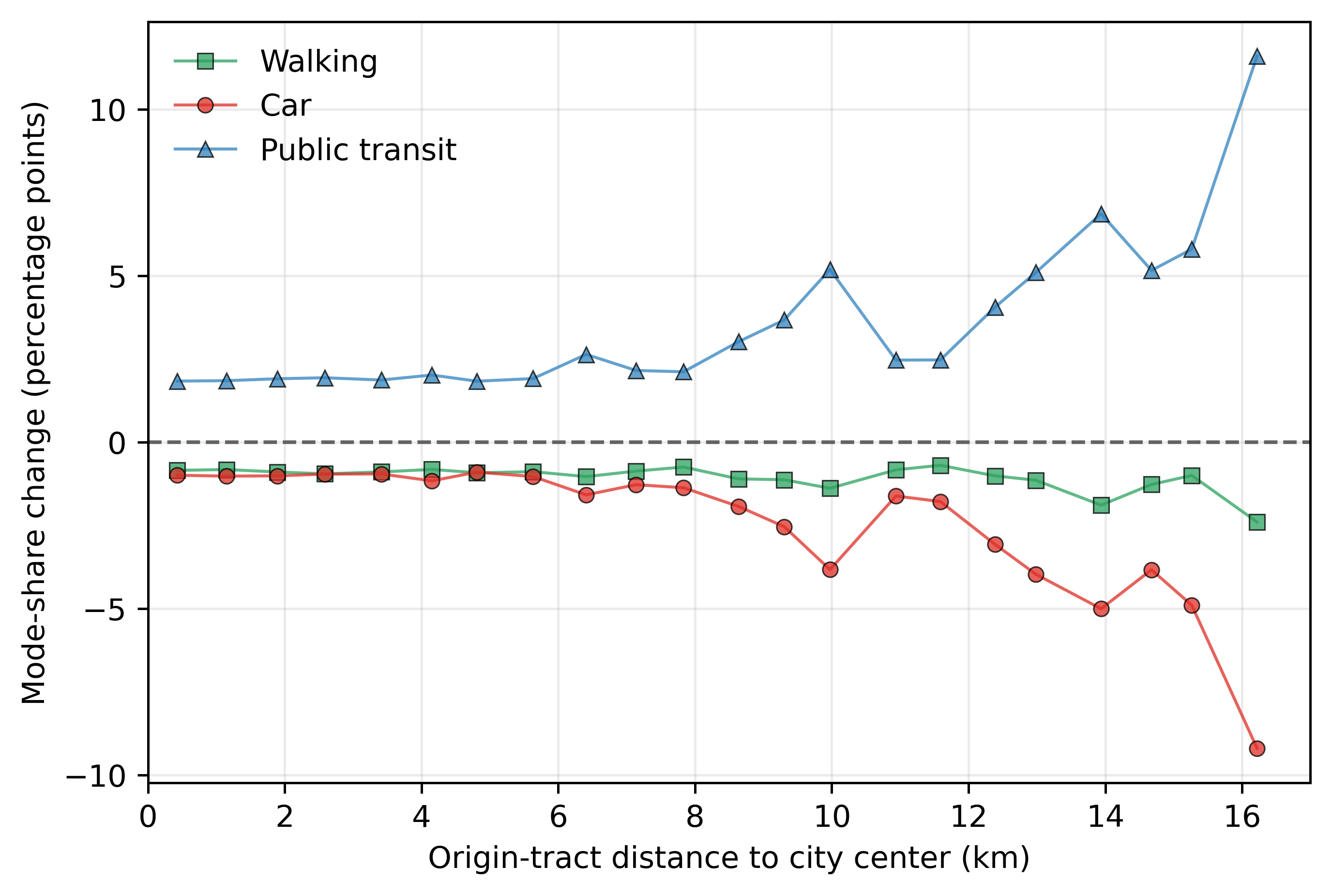}
    \includegraphics[width=0.545\linewidth]{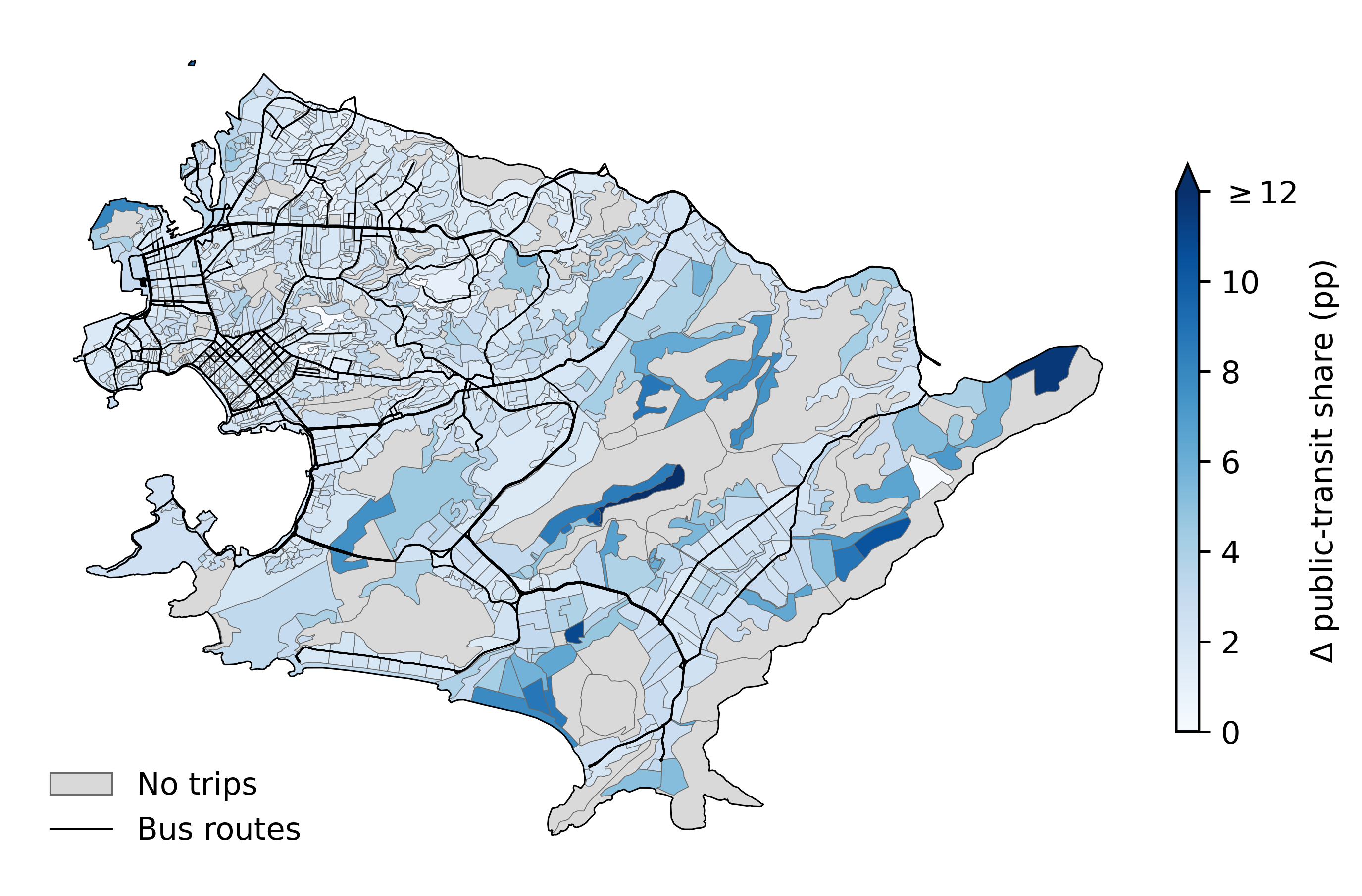}
    \caption{\textbf{Counterfactual scenario increases public transit usage in Niter\'oi.} The counterfactual scenario reduces by 30\% the time required to travel between trip origins or destinations and bus stops, while retaining the same trips and other model settings as in the baseline simulation. The left panel reports the scenario-minus-baseline change in modal share, in percentage points, for \SI{750}{\metre} intervals of origin-tract distance from the city center. 
    Public-transit share increases in every distance interval: gains are approximately two percentage points across much of the inner city and generally become larger beyond \SI{10}{\kilo\metre}. These gains are offset primarily by reductions in car use and, to a lesser extent, walking. The right panel maps the change in public-transit share by trip-origin census tract. Blue indicates an increase and red a decrease relative to the baseline, with darker colors representing larger changes; gray tracts contain no trips that could be paired between the baseline and counterfactual simulations, and black lines show the bus-route network. The widespread positive response indicates that shorter access and egress times can increase the attractiveness of public transit across the city.}
    \label{fig:scenario_test}
\end{figure}

At the city scale, reducing public-transit access and egress times by
30\% increases the modeled public-transit-to-car trip ratio from 1.843
to 2.052, a relative increase of 11.3\%. Thus, the model assigns about
2.05 public-transit trips for every car trip in the scenario, compared
with 1.84 in the baseline, while trip demand, destinations, and preference
parameters remain unchanged. This ratio change summarizes the shift
toward transit relative to car travel; it is distinct from the
percentage-point changes in modal shares shown in
Fig.~\ref{fig:scenario_test}.

The citywide shift is accompanied by a clear distance gradient
(Fig.~\ref{fig:scenario_test}, left). Public-transit share increases
across all distance bands, with gains of approximately 2 percentage
points through much of the inner city and larger increases beyond
about 10~km from the city center. Gains exceed 5 percentage points in
the outermost bands. These increases are offset primarily by reductions
in car share and, to a lesser extent, walking share.

The tract-level map reveals local variation within this broad gradient
(Fig.~\ref{fig:scenario_test}, right). Most tracts show increased transit
share, but the gains are uneven and do not simply trace the bus-route
network. Under the model, the response depends on the absolute access
and egress time saved and the baseline probabilities of the competing
modes. Route presence alone therefore does not determine the magnitude
of the modeled change.

Comparing these changes with the baseline mobility classes
(Fig.~\ref{fig:spatial_patterns_transmode}) helps identify where the
scenario shifts the balance between modes. The walking-oriented western
areas generally show modest transit gains, while many larger gains occur
in peripheral car-oriented tracts or near the boundaries between car-
and transit-oriented areas. Several tracts with already high transit
shares also gain. This spatial correspondence suggests that improved
access can increase transit's modeled competitiveness in car-oriented
areas while also reinforcing its role in existing transit-oriented areas.



Overall, in this ``what-if'' scenario the modeled public-transit-to-car
ratio increases by 11.3\%, with larger transit-share gains in several peripheral,
car-oriented areas. The key finding is the uneven spatial response.
The same proportional improvement in access produces different changes
in transportation choice across the city. This identifies areas where
first- and last-mile improvements warrant closer investigation.

\section{Conclusion}
\label{sec:conclusion}

This study shows how macroscopic and mostly static urban data can support
a microscopic, dynamic, and spatially detailed representation of mobility
in Niter\'oi. Building on the patterns-of-life simulator developed by
Reia et al.~\citep{reia2026}, we integrate activity-generated trips with
road and bus-network travel times and probabilistic mode assignment.
This integration connects urban population and infrastructure data to
geographic differences in transportation choice and enables controlled
experiments with transit accessibility.

The simulated patterns exhibit a broad center--periphery gradient.
Walking is more prevalent in compact central areas, public transit is
important at intermediate distances and along bus corridors, and car
use increases toward the periphery. Local variation within this gradient
is consistent with differences in activity locations and transport
access. Simulated neighborhood-level transit activity is strongly
associated with the empirical proxy, with Pearson $r=0.885$ and
Spearman $\rho=0.873$. This supports correspondence in aggregate transit
geography, although the shared route-based spatial allocation may
contribute to the association. The comparison does not independently
validate individual travel decisions or walking and car shares.

The transit-access scenario illustrates how the framework can identify
spatial differences in the modeled response to an intervention.
Reducing walking times to and from bus stops by 30\% increases the
modeled public-transit-to-car trip ratio by 11.3\%, while holding trip
demand, destinations, and preference parameters fixed. Transit-share
gains are larger in several peripheral areas, including car-oriented
tracts, than in much of the inner city. The main finding is this uneven
response to the same proportional improvement in access, which identifies
areas where first- and last-mile measures warrant closer investigation.

These findings are conditional on the available data and modeling
assumptions. Activity locations and destination choices depend on
simplified building classifications and behavioral rules, while mode
assignment omits household constraints, vehicle ownership, monetary
costs, and individual preference differences. Calibration uses regional
modal shares rather than locally observed choices, and the intervention
response depends on the assumed travel-time coefficient, constant
speeds, and simplified transit routing and service representation.
Evaluation is limited to a route-weighted transit-activity proxy.
The uniform access-time reduction also represents a hypothetical
improvement rather than a specific project and does not establish
implementation feasibility, costs, capacity effects, or induced demand.
Sensitivity analysis and additional local observations are therefore
needed before interpreting the scenario magnitudes as forecasts.

The framework connects static urban data to spatially differentiated
mobility scenarios, providing a basis for identifying where more detailed
observation and transit-access planning would be most useful. Its value
in cities with limited travel data lies in making the links between
activity demand, transport access, and modeled transportation choices
explicit, so that potential improvements can be examined in their
geographic context.

\backmatter

\bmhead{Supplementary information}

Supplementary figures and supporting analyses are provided in the accompanying file \texttt{supplementary\_material.pdf}.


\section*{Declarations}

\bmhead{Funding}

This work was supported by the Intelligence Advanced Research Projects Activity (IARPA) via Department of Interior/Interior Business Center (DOI/IBC) contract number 140D0419C0050. The U.S. Government is authorized to reproduce and distribute reprints for Governmental purposes notwithstanding any copyright annotation thereon. Disclaimer: The views and conclusions contained herein are those of the authors and should not be interpreted as necessarily representing the official policies or endorsements, either expressed or implied, of IARPA, DOI/IBC, or the U.S. Government.
This work was supported in part by the National Science Foundation award number 2109647.

\bmhead{Competing interests}

The authors declare no competing financial or non-financial interests

\bmhead{Ethics approval and consent to participate}

Not applicable.

\bmhead{Consent for publication}

Not applicable.

\bmhead{Data availability}

All empirical datasets used in this study are publicly available through their original providers. The data sources, their respective applications, and links for accessing the datasets are provided in Table \ref{tab:data}. Derived datasets generated during this study, including synthetic populations and simulated mobility records, are available from the corresponding author upon reasonable request.

\bmhead{Materials availability}

Not applicable.

\bmhead{Code availability}

The computational codes used to process the data and perform the analyses presented in this study will be made available by the corresponding author upon reasonable request.

\bmhead{Author contributions}

S.M.R., T.A., and D.P. conceived the study and contributed to the methodology. T.A. and D.P. acquired funding, while D.P. provided resources and administered the project. S.M.R. curated the data, performed the formal analysis and validation, and prepared the visualizations. S.M.R., T.A., and D.P. contributed to the investigation and preparation of the original manuscript. All authors reviewed and edited the manuscript and approved its final version.

\bibliography{references}

@book{benakiva1985discrete,
  title     = {Discrete Choice Analysis: Theory and Application to Travel Demand},
  author    = {Ben-Akiva, Moshe and Lerman, Steven R.},
  publisher = {MIT Press},
  address   = {Cambridge, MA},
  year      = {1985}
}

@misc{ibge2022census,
  author       = {{Instituto Brasileiro de Geografia e Estat\'istica}},
  title        = {Censo Demogr\'afico 2022},
  year         = {2022},
  howpublished = {IBGE},
  url          = {https://www.ibge.gov.br/estatisticas/sociais/populacao/22827-censo-demografico-2022.html},
  note         = {Accessed 1 September 2026}
}

@incollection{mcfadden1974conditional,
  title     = {Conditional logit analysis of qualitative choice behavior},
  author    = {McFadden, Daniel},
  booktitle = {Frontiers in Econometrics},
  editor    = {Zarembka, Paul},
  pages     = {105--142},
  publisher = {Academic Press},
  address   = {New York},
  year      = {1974}
}

@book{train2009discrete,
  title     = {Discrete Choice Methods with Simulation},
  author    = {Train, Kenneth E.},
  edition   = {2},
  publisher = {Cambridge University Press},
  address   = {Cambridge},
  year      = {2009},
  doi       = {10.1017/CBO9780511805271}
}

@article{barandier2015backcasting,
  title   = {Applying the backcasting method to achieve sustainable mobility: The case of Niter\'oi},
  author  = {Barandier, Jr., Jos\'e Renato},
  journal = {Transportation Research Procedia},
  volume  = {8},
  pages   = {5--16},
  year    = {2015},
  doi     = {10.1016/j.trpro.2015.06.037}
}

@article{grieco2010trip,
  title   = {Taxas de gera\c{c}\~ao de viagens em condom\'inios residenciais de Niter\'oi: Estudo de caso},
  author  = {Grieco, Elisabeth Poubel and Portugal, Licinio da Silva},
  journal = {Transportes},
  volume  = {18},
  number  = {1},
  year    = {2010},
  doi     = {10.14295/transportes.v18i1.386}
}

@article{grieco2020built,
  title   = {Influence of the built environment on sustainable trips: The case of Niter\'oi, Brazil},
  author  = {Grieco, Elisabeth and da Silva Portugal, Licinio and Alves, Rosane},
  journal = {Proceedings of the Institution of Civil Engineers -- Municipal Engineer},
  volume  = {173},
  number  = {1},
  pages   = {14--24},
  year    = {2020},
  doi     = {10.1680/jmuen.17.00010}
}

@article{reia2025function,
  title   = {Function and form of {US} cities},
  author  = {Reia, Sandro M. and Anderson, Taylor and de Arruda, Henrique F. and Atwal, Kuldip S. and Ruan, Shiyang and Kavak, Hamdi and Pfoser, Dieter},
  journal = {Computers, Environment and Urban Systems},
  volume  = {116},
  pages   = {102221},
  year    = {2025},
  doi     = {10.1016/j.compenvurbsys.2024.102221}
}

@article{zufle2023urban,
  title   = {Urban life: A model of people and places},
  author  = {Z\"ufle, Andreas and Wenk, Carola and Pfoser, Dieter and Crooks, Andrew and Kim, Joon-Seok and Kavak, Hamdi and Manzoor, Umar and Jin, Hyunjee},
  journal = {Computational and Mathematical Organization Theory},
  volume  = {29},
  number  = {1},
  pages   = {20--51},
  year    = {2023},
  doi     = {10.1007/s10588-021-09348-7}
}

@article{primerano2008defining,
  title   = {Defining and understanding trip chaining behaviour},
  author  = {Primerano, Frank and Taylor, Michael A. P. and Pitaksringkarn, Ladda and Tisato, Peter},
  journal = {Transportation},
  volume  = {35},
  number  = {1},
  pages   = {55--72},
  year    = {2008},
  doi     = {10.1007/s11116-007-9134-8}
}

@article{gonzalez2008,
  author  = {Gonz{\'a}lez, Marta C. and Hidalgo, C{\'e}sar A. and Barab{\'a}si, Albert-L{\'a}szl{\'o}},
  title   = {Understanding individual human mobility patterns},
  journal = {Nature},
  year    = {2008},
  volume  = {453},
  pages   = {779--782},
  doi     = {10.1038/nature06958}
}

@article{song2010scaling,
  author  = {Song, Chaoming and Koren, Tal and Wang, Pu and Barab{\'a}si, Albert-L{\'a}szl{\'o}},
  title   = {Modelling the scaling properties of human mobility},
  journal = {Nature Physics},
  year    = {2010},
  volume  = {6},
  pages   = {818--823},
  doi     = {10.1038/nphys1760}
}

@article{song2010predictability,
  author  = {Song, Chaoming and Qu, Zehui and Blumm, Nicholas and Barab{\'a}si, Albert-L{\'a}szl{\'o}},
  title   = {Limits of predictability in human mobility},
  journal = {Science},
  year    = {2010},
  volume  = {327},
  pages   = {1018--1021},
  doi     = {10.1126/science.1177170}
}

@article{schneider2013,
  author  = {Schneider, Christian M. and others},
  title   = {Unravelling daily human mobility motifs},
  journal = {Journal of the Royal Society Interface},
  year    = {2013},
  doi     = {10.1098/rsif.2013.0246}
}

@article{demontjoye2013,
  author  = {de Montjoye, Yves-Alexandre and Hidalgo, C{\'e}sar A. and Verleysen, Michel and Blondel, Vincent D.},
  title   = {Unique in the crowd: The privacy bounds of human mobility},
  journal = {Scientific Reports},
  year    = {2013},
  volume  = {3},
  pages   = {1376},
  doi     = {10.1038/srep01376}
}

@article{chang2021,
  author  = {Chang, Serina and others},
  title   = {Mobility network models of {COVID-19} explain inequities and inform reopening},
  journal = {Nature},
  year    = {2021},
  volume  = {589},
  pages   = {82--87},
  doi     = {10.1038/s41586-020-2923-3}
}

@article{cervero1997,
  author  = {Cervero, Robert and Kockelman, Kara},
  title   = {Travel demand and the 3{D}s: Density, diversity, and design},
  journal = {Transportation Research Part D: Transport and Environment},
  year    = {1997},
  volume  = {2},
  number  = {3},
  pages   = {199--219},
  doi     = {10.1016/S1361-9209(97)00009-6}
}

@article{ewing2010,
  author  = {Ewing, Reid and Cervero, Robert},
  title   = {Travel and the built environment: A meta-analysis},
  journal = {Journal of the American Planning Association},
  year    = {2010},
  volume  = {76},
  number  = {3},
  doi     = {10.1080/01944361003766766}
}

@article{geurs2004,
  author  = {Geurs, Karst T. and van Wee, Bert},
  title   = {Accessibility evaluation of land-use and transport strategies: Review and research directions},
  journal = {Journal of Transport Geography},
  year    = {2004},
  volume  = {12},
  pages   = {127--140},
  doi     = {10.1016/j.jtrangeo.2003.10.005}
}

@article{beckman1996,
  author  = {Beckman, Richard J. and Baggerly, Keith A. and McKay, Michael D.},
  title   = {Creating synthetic baseline populations},
  journal = {Transportation Research Part A: Policy and Practice},
  year    = {1996},
  volume  = {30},
  number  = {6},
  pages   = {415--429},
  doi     = {10.1016/0965-8564(96)00004-3}
}

@inproceedings{ye2009,
  author    = {Ye, Xin and Konduri, Karthik and Pendyala, Ram M. and Sana, Bhargava and Waddell, Paul},
  title     = {A methodology to match distributions of both household and person attributes in the generation of synthetic populations},
  booktitle = {Proceedings of the 88th Annual Meeting of the Transportation Research Board},
  year      = {2009},
  url       = {https://trid.trb.org/View/881554}
}

@article{farooq2013,
  author  = {Farooq, Bilal and Bierlaire, Michel and Hurtubia, Ricardo and Fl{\"o}tter{\"o}d, Gunnar},
  title   = {Simulation based population synthesis},
  journal = {Transportation Research Part B: Methodological},
  year    = {2013},
  volume  = {58},
  pages   = {243--263},
  doi     = {10.1016/j.trb.2013.09.012}
}

@article{horl2021,
  author  = {H{\"o}rl, Sebastian and Balac, Milo{\v{s}}},
  title   = {Synthetic population and travel demand for {Paris} and {{\^I}le-de-France} based on open and publicly available data},
  journal = {Transportation Research Part C: Emerging Technologies},
  year    = {2021},
  doi     = {10.1016/j.trc.2021.103291}
}

@article{boeing2017,
  author  = {Boeing, Geoff},
  title   = {{OSMnx}: New methods for acquiring, constructing, analyzing, and visualizing complex street networks},
  journal = {Computers, Environment and Urban Systems},
  year    = {2017},
  doi     = {10.1016/j.compenvurbsys.2017.05.004}
}

@article{bowman2001,
  author  = {Bowman, John L. and Ben-Akiva, Moshe E.},
  title   = {Activity-based disaggregate travel demand model system with activity schedules},
  journal = {Transportation Research Part A: Policy and Practice},
  year    = {2001},
  volume  = {35},
  number  = {1},
  pages   = {1--28},
  doi     = {10.1016/S0965-8564(99)00043-9}
}

@article{arentze2000,
  author  = {Arentze, Theo and Hofman, Frank and van Mourik, Henk and Timmermans, Harry},
  title   = {{ALBATROSS}: Multiagent, rule-based model of activity pattern decisions},
  journal = {Transportation Research Record},
  year    = {2000},
  doi     = {10.3141/1706-16}
}

@article{bhat2004,
  author  = {Bhat, Chandra R. and Guo, Jessica Y. and Srinivasan, Sivaramakrishnan and Sivakumar, Aruna},
  title   = {Comprehensive econometric microsimulator for daily activity-travel patterns},
  journal = {Transportation Research Record},
  year    = {2004},
  volume  = {1894},
  doi     = {10.3141/1894-07}
}

@incollection{mcfadden1974,
  author    = {McFadden, Daniel},
  title     = {Conditional logit analysis of qualitative choice behavior},
  booktitle = {Frontiers in Econometrics},
  editor    = {Zarembka, Paul},
  publisher = {Academic Press},
  year      = {1974},
  pages     = {105--142},
  url       = {https://eml.berkeley.edu/~mcfadden/iatbr00.html}
}

@book{benakiva1985,
  author    = {Ben-Akiva, Moshe E. and Lerman, Steven R.},
  title     = {Discrete Choice Analysis: Theory and Application to Travel Demand},
  publisher = {MIT Press},
  year      = {1985},
  url       = {https://mitpress.mit.edu/9780262022170/discrete-choice-analysis/}
}

@misc{reia2026,
  author        = {Reia, Sandro M. and de Arruda, Henrique F. and Ruan, Shiyang and Anderson, Taylor and Kavak, Hamdi and Pfoser, Dieter},
  title         = {Towards Universal Urban Patterns-of-Life Simulation},
  year          = {2026},
  eprint        = {2601.22099},
  archivePrefix = {arXiv},
  url           = {https://arxiv.org/abs/2601.22099}
}

@article{drchal2016,
  author  = {Drchal, Jan and {\v{C}}ertick{\'y}, Michal and Jakob, Michal},
  title   = {{VALFRAM}: Validation framework for activity-based models},
  journal = {Journal of Artificial Societies and Social Simulation},
  year    = {2016},
  volume  = {19},
  number  = {3},
  pages   = {5},
  doi     = {10.18564/jasss.3127}
}

@article{stopher2007,
  author  = {Stopher, Peter R. and Greaves, Stephen P.},
  title   = {Household travel surveys: Where are we going?},
  journal = {Transportation Research Part A: Policy and Practice},
  year    = {2007},
  volume  = {41},
  number  = {5},
  pages   = {367--381},
  doi     = {10.1016/j.tra.2006.09.005}
}

@article{shen2014,
  author  = {Shen, Li and Stopher, Peter R.},
  title   = {Review of {GPS} travel survey and {GPS} data-processing methods},
  journal = {Transport Reviews},
  year    = {2014},
  volume  = {34},
  number  = {3},
  pages   = {316--334},
  doi     = {10.1080/01441647.2014.903530}
}

@misc{ibge2022,
  author       = {{Instituto Brasileiro de Geografia e Estat{\'i}stica}},
  title        = {2022 Population Census},
  year         = {2022},
  howpublished = {\url{https://www.ibge.gov.br/en/statistics/social/labor/22836-2022-census-3.html}},
  note         = {Population by age and sex and geographic census results}
}

@article{klinkhardt2021,
  author  = {Klinkhardt, Christian and W{\"o}rle, Tim and Briem, Lars
             and Heilig, Michael and Kagerbauer, Martin and Vortisch, Peter},
  title   = {Using {OpenStreetMap} as a Data Source for Attractiveness
             in Travel Demand Models},
  journal = {Transportation Research Record},
  year    = {2021},
  doi     = {10.1177/0361198121997415}
}

@misc{vida2024openbuildings,
  author       = {{VIDA}},
  title        = {{Google-Microsoft-OSM Open Buildings: Combined by VIDA}},
  year         = {2024},
  version      = {2.0},
  publisher    = {Source Cooperative},
  howpublished = {Brazilian S2-partitioned GeoParquet collection},
  url          = {https://source.coop/vida/google-microsoft-osm-open-buildings/geoparquet/by_country_s2/country_iso=BRA},
  note         = {Dataset combining Google V3 Open Buildings, Microsoft GlobalML Building Footprints, and OpenStreetMap building footprints; accessed 29 September 2026}
}

@misc{sigeoniteroi,
  author       = {{Prefeitura Municipal de Niterói}},
  title        = {{HUB SIGeo: Sistema de Gestão da Geoinformação de Niterói}},
  year         = {n.d.},
  publisher    = {Prefeitura Municipal de Niterói},
  howpublished = {Municipal geographic data portal},
  url          = {https://www.sigeo.niteroi.rj.gov.br},
  note         = {Source of the Niterói census-sector and municipal facility GeoJSON layers; accessed 29 September 2026}
}

@techreport{tumi2020niteroi,
  author      = {{Transformative Urban Mobility Initiative}},
  title       = {Municipality of Niter\'oi, Brazil:
                 Creating Momentum Towards a Walkable and Bikeable City},
  institution = {Transformative Urban Mobility Initiative},
  year        = {2020},
  month       = apr,
  url         = {https://transformative-mobility.org/wp-content/uploads/2023/03/TUMI_City-profile-and-story-Niteroi-QjKjKD.pdf}
}

@misc{sigeorotasmobnit,
  author       = {{Prefeitura Municipal de Niter\'oi}},
  title        = {{RotasMobNit}: Bus Route Geometries},
  howpublished = {SIGeo ArcGIS REST service},
  url          = {https://sig.niteroi.rj.gov.br/server/rest/services/Hosted/RotasMobNit/FeatureServer/0},
  note         = {Route geometries and line attributes; layer 0}
}

@misc{sigeoparadas,
  author       = {{Prefeitura Municipal de Niter\'oi}},
  title        = {Paradas de \^Onibus},
  howpublished = {SIGeo ArcGIS REST service},
  url          = {https://sig.niteroi.rj.gov.br/server/rest/services/Hosted/Paradas_de_%C3%94nibus/FeatureServer},
  note         = {Bus-stop locations (layer 0), bus-line catalog
                  (table 3), and stop--line associations (table 4)}
}

@book{zipf1949,
  author    = {Zipf, George Kingsley},
  title     = {Human Behavior and the Principle of Least Effort:
               An Introduction to Human Ecology},
  publisher = {Addison-Wesley Press},
  address   = {Cambridge, MA},
  year      = {1949}
}

@article{alessandretti2018conserved,
  author  = {Alessandretti, Laura and Sapiezynski, Piotr
             and Sekara, Vedran and Lehmann, Sune
             and Baronchelli, Andrea},
  title   = {Evidence for a Conserved Quantity in Human Mobility},
  journal = {Nature Human Behaviour},
  volume  = {2},
  number  = {7},
  pages   = {485--491},
  year    = {2018},
  doi     = {10.1038/s41562-018-0364-x},
  url     = {https://doi.org/10.1038/s41562-018-0364-x}
}

@misc{openstreetmap,
  author       = {{OpenStreetMap contributors}},
  title        = {{OpenStreetMap}},
  howpublished = {Geographic database},
  url          = {https://www.openstreetmap.org/copyright},
  note         = {Road network data for Niter\'oi, Brazil.
                  Available under the Open Database License (ODbL)}
}

@techreport{PMUSNiteroi2020,
  author      = {{Secretaria Municipal de Urbanismo e Mobilidade de Niter{\'o}i}},
  title       = {Plano de Mobilidade Urbana Sustent{\'a}vel de Niter{\'o}i: PMUS Niter{\'o}i 2020--2030},
  institution = {Prefeitura Municipal de Niter{\'o}i},
  type        = {Relat{\'o}rio Final},
  address     = {Niter{\'o}i, Rio de Janeiro, Brazil},
  year        = {2020},
  url         = {https://urbanismo.niteroi.rj.gov.br/pmus.html},
  note        = {Section 6.1, Mobilidade Urbana de Niter{\'o}i em n{\'u}meros}
}

\end{document}